\documentclass[aps,superscriptaddress,twocolumn,preprintnumbers,showkeys,showpacs,amsmath,amssymb]{revtex4}
\usepackage[cp1251]{inputenc}
\usepackage[english]{babel}
\usepackage{amssymb,amsmath}
\usepackage{amstext,textcomp} 
\usepackage[dvips]{hyperref}
\usepackage[mathscr]{eucal}
\usepackage{longtable}
\usepackage{graphicx}
\begin{document}
\title{To Principles of Quantum Mechanics Development}
\author{Dmitri Yerchuck (a),  Alla  Dovlatova (b), Felix Borovik (a) Yauhen Yerchak  (c), Vyacheslav Stelmakh (c)\\
\textit{(a) Heat-Mass Transfer Institute of National Academy of Sciences of RB, Brovka Str.15, Minsk, 220072,\\ 
(b) M.V.Lomonosov Moscow State University, Moscow, 119899, RF \\(c)  Belarusian State University, Nezavisimosti Ave., 4, Minsk, 220030, RB}}
\date{\today}%

\begin{abstract} New insight to the  principles of the quantum physics development  is given.  

 The correct ways for the construction of  new versions of quantum mechanics on the second main postulate base are discussed. 

The conclusion on the status of the second main postulate is given. Its formulation in all textbooks has to be represented in the form of statement, since the hypothesis of  Schr\"{o}dinger on the existance of the field scalar function, being to be observable quantity, just charge density, is strictly proved for the case of EM-field, the role of which is argued to be decisive for the dynamics of the atomic systems. It is shown, that  the field scalar function, being to be the function the only of coordinates and time, actually describes the state of the system.
 The second main postulate in Schr\"{o}dinger formulation is mathematically strictly grounded, but in the  popular probabilistic form used in modern textbooks on quantum theory it cannot be proved. The probabilistic theatise, proposed by Born  is true in a number of special cases, quite correctly indicated by Dirac. 

The possible ways of the development of quantum theory, based on clear understanding of the origin of corpuscular-wave dualism  are analysed.

\end{abstract}  

\maketitle \section{Introduction and Background} 

The progress in  the modern technology branches - nanoelectronics, spintronics, developing on the use of various quantum physics phenomena, the successes in the  elaboration of logic quantum systems including quantum
computers and quantum communication systems require the corresponding development of the quantum theory. To develop the quantum theory it seems to be significant  to know, whether is the sufficiently  firm its  base. In other words, it is required to establish, whether can be strictly grounded the main postulates of the quantum theory (or not)and a rather deep insight into the principles of its construction at all.
 The first main postulate, from which is begun  the study of quantum theory in all textbooks (see, for example, \cite{Landau_L.D}, \cite{Kramers}, \cite{Bohm}, \cite{Davydov}) is the following: "The linear self-adjoint operators are setting up into correspondence to observable physical quantities". Its fundamentality was recently analysed in the work \cite{Dmitri_Alla_Andrey}. It has been shown, that the given postulate can be mathematically strictly proved. The proof is based on the symmetry study of main differential equations of mechanics and electrodynamics. It   has been shown, that differential equations, which are invariant under transformations of groups, which are symmetry groups of mathematical numbers (considered within the frames of the number theory) determine the mathematical nature of the
quantities, incoming in given equations. So, it was strictly proved that mechanical quantiies are complex-valued quantities. Let us give some details. It seems to be substantial that differential
equations for the dynamics of the  nonrelativistic classical mechanics
 are  invariant under
transformation 
\begin{equation}
\label{eq10w}
u'(x) = \beta exp(i \alpha) u(x),
\end{equation}
where $\alpha,\beta\in R$,  $u(x)$ is the set  of corresponding mechanical variables. Really, the dynamics of classical mechanics systems is described by Lagrange equations or by equivalent to them canonical Hamilton equations. It is known, that, for instance, Hamilton equations are invariant under contact transformations of variables, that is, under the transformation of linear elements - positions and directions, but not the  points' transformation. The transformation $(\ref{eq10w})$ is referred to the  contact transformations' class. It means, that  all physical
quantities, which determine  the dynamics of classical mechanics systems  have to be represented by  a quantum-mechanical description
 by one of the variants of the
representation of complex numbers. In its turn, it has been shown in the works \cite{A_Dovlatova_D_Yerchuck} and \cite{Dmitri_Alla_Andrey} that there is  existing the infinite number of the variants of the representation of complex numbers. Along with the basis of the linear space of complex numbers over the field $R$ of real numbers  being to be consiting  of the numbers $1$ and $i$ to any complex number $a + ib$, $a, b \in R$, represented in the given basis, can be set up in conformity the $[2 \times 2]$-matrix according to biective mapping

\begin{equation}
\label{eq1abcd}
 f : a + ib \to \left[\begin{array} {*{20}c} a&-b  \\ b&a \end{array}\right].
\end{equation}

From the bijectivity of mapping (\ref{eq1abcd}) follows the existence of  inverse mapping. It means that  to any matrix, which has the structure, given by the right side in the relation (\ref{eq1abcd}), corresponds the complex number, determined by the left side. 
 The matrices
\begin{equation}
\label{eq61}
\left[\begin{array} {*{20}c} 1&0  \\ 0&1 \end{array}\right], 
\left[\begin{array} {*{20}c} 0&-i  \\ i&0 \end{array}\right]
\end{equation}
produce  the basis for complex numbers $\{a + ib\}$, $a,b \in R$  in the linear space  of $[2 \times 2]$-matrices, defined over the field of real numbers.

 It is argued in above cited work, that it is convenient often to define the space of complex numbers over the group $Z$ of real positive numbers including zero. In the given case the  dimensionality of the matrices and  their basis has to be duplicated, since to two unities - positive $1$ and negative $-1$ 
can be set up in conformity the $[2 \times 2]$-matrices according to the following biective mapping
\begin{equation}
\label{eq62}
 \xi : 1 \to \left[\begin{array} {*{20}c} 1&0  \\ 0&1  \end{array}\right], -1 \to \left[\begin{array} {*{20}c} 0&1  \\ 1&0 \end{array}\right].
\end{equation}
  So, the following $[4 \times 4]$-matrices, so called [0,1]-matrices, 
\begin{equation}
\label{eq63}
\begin{split}
\raisetag{40pt}
&\zeta : 1 \to [e_1]=\left[\begin{array} {*{20}c}1&0&0&0 \\  0&1&0&0  \\ 0&0&1&0 \\  0&0&0&1 \end{array}\right], \\
&i \to  [e_2]=\left[\begin{array} {*{20}c}  0&1&0&0  \\ 0&0&1&0 \\0&0&0&1\\ 1&0&0&0 \end{array}\right],\\
&-1 \to [e_3]=\left[\begin{array} {*{20}c} 0&0&1&0 \\0&0&0&1\\ 1&0&0&0\\0&1&0&0  \end{array}\right],\\
&-i \to  [e_4]=\left[\begin{array} {*{20}c}  0&0&0&1\\ 1&0&0&0\\0&1&0&0  \\ 0&0&1&0 \end{array}\right].
\end{split}
\end{equation}
can also be  the basis of the space of complex numbers over the group $Z$ of real positive numbers (including zero).

We see from the given example, that the choise of the basis is ambiguous. Any four  $[4 \times 4]$  [0,1]-matrices, which satisfy the rules of the cyclic recurrence
\begin{equation}
\label{eq64}
i^1 = i, i^2 = -1, i^3= -i, i^4 = 1
\end{equation}
 can be the basis  of the space of complex numbers over the same group $Z$ of real positive numbers (including zero).

Thus, we have elucidated the conclusion in \cite{A_Dovlatova_D_Yerchuck} and \cite{Dmitri_Alla_Andrey}, that the system of complex numbers can be constructed by the infinite number of the ways, at that, the cyclic basis can consist of $m$ units, $ m \in N$, starting from three. 

Let us accentuate once again, that it is remarkable, that the conformity between complex numbers and matrices is realized by biective mappings.  Conseqyently, to any   squarte matrix, belonging to the linear space with a basis given by (\ref{eq63}),  or any other, satisfying the rules of the cyclic recurrence  (\ref{eq64}) can be set up in conformity like to mapping (\ref{eq1abcd}) the complex number. 

For the applications in quantum physics is especially significant, that to any Hermitian  matrix $H$ can be set up in conformity the following complex number
\begin{equation}
\label{eq66}
  \zeta: H \to  S + iA = \left[\begin{array} {*{20}c} S&-A  \\ A&S  \end{array}\right] 
\end{equation}
and vice versa. Here $S$  and $A$  are symmetric and antisymmetric parts of a Hermitian  matrix.

Just in the given posssibility to represent the complex numbers in the form of the Hermitian matrices and in the  invariance of the equations of mechanical dynamics under the transformations of the multiplicative group of the complex number field consists the proof of the
statement, being to be the proof, obtained in \cite{Dmitri_Alla_Andrey}, of the
main postulate of quantum mechanics:

\textit{To any mechanical quantity can be set
up in the correspondence the Hermitian
matrix by  the quantization.}

The choose of the construction of the mathemaical apparatus of quantum mechanics on the base of Hermitian matrices is convenient,
however, it is the only one variant from the
infinity of variants of the representations
of quantum mechanical quantities by complex numbers.
It was rightly concluded in \cite{Dmitri_Alla_Andrey}, that the given results open the high functional  possibilities for the development of the quantum theory.

It is especially interesting, that the possibility to represent the Hermitian matrices by complex numbers was known to Dirac. Moreover, he used the given transformations in his calculations, see for example \cite{Dirac_P_A_M}.  So, Dirac was very near to prove the first main postulate.

It was also shown, that a non-abelian character of the  multiplicative group of the quaternion ring leads to the nonapplicability of the quaternion calculus for the construction of  new versions of quantum mechanics directly. In other words, mechanical quantities are not quaternions. They are in general case also not real numbers. The description of mechanical dynamical systems within the limits of the field of real numbers will be correct the only in the case when the phase shifts between the corresponding mechanical variables tend to zero. It is well known, that it is the case of classical Newton mechanics systems. 

In contrast to mechanical quantities, the electrodynamics quantities are quaternions. The given conclusion follows from the proof of the invariance of Maxwell equations  under transformations of the quaternion non-abelian multiplicative group, obtained in in \cite{A_Dovlatova_D_Yerchuck} and \cite{Dmitri_Alla_Andrey}. Consequently  to any electrodynamics quantity can
be set up in the correspondence the
quaternion matrix  by the quantization of EM-field. It is in fact the consequence of the presence along with the symmetry of Maxwell equations under transformations, given by  $(\ref{eq10w})$, the Rainich \cite{Rainich} dual symmetry and the additional dual symmetry, established in \cite{A_Dovlatova_D_Yerchuck}.
We can wonder to Maxwell deep insight in the field, and to his intuition in the correct definition of electrodynamics quantities to be quaternions. However, the quaternion content was made vapid from the electrodynamics theory in the subsequent studies, owing mainly to  the works of Heaviside \cite{Heaviside}. It was recovered the only in \cite{A_Dovlatova_D_Yerchuck}, that is, more than a hundred years after Heaviside works. Let us remark, that the proof of the quaternionic nature of electromagnetic quantities in \cite{A_Dovlatova_D_Yerchuck}, \cite{Dmitri_Alla_Andrey} is in fact the proof, that full set of basic functions for the description of electromagnetic phenomena has to be consisting from four scalar functions, or equivalently, from one vector and one scalar functions, that really is used in the practice of the solution of the electrodynamics tasks.

We wish also remark, that the   conclusion on the nonapplicability of the quaternion calculus for the description of mechanical systems
 seems to be actual, since there is a number of modern publications with the development of the quantum mechanics theory using the quaternions directly with the standard basis \{e, i, j, k \}.
 The correct way for the construction of  new versions of quantum mechanics on the quaternion base is however in the pincple possible, however, it is indirect.  It was discussed in the paper \cite{DY_DA_A}, where it has been shown, that the correct way can be realized by means of the representation  of the quaternions through the basis of the linear space of complex numbers over the field of real numbers, under the multiplicative group  of which the equations of the dynamics of mechanical systems are  invariant.

 Summing up  the premises, we see, that the first main postulate, being to be now stricly mathematically grounded, is really the corner stone, giving the firm base for the development of the quantum theory. It is  the consequence of more general laws of the Nature, just  the symmetry of mechanics and electrodynamics equations which was not taken into account by founding fathers of the quantum theory and their followers.

\section{Analysis of the Second Main Postulate and the Ways of Quantum Theory Development}

It represents the interest to establish, wheher the second main postulate is also
the corner stone, giving the firm base for the development of the quantum theory or not.  It is the main task of the work presented.
 The second main postulate is formulated in texbooks on quantum mechanics in the following way, in particular, in \cite{Landau_L.D}:"The state of a system can be described by a certain (generally speaking complex) function of coordinates $\Psi(q)$, at that the square of the module of this function determines the probability distribution of the coordinates' values: $|\Psi(q)|^2dq$ is a probability of that, that a measurement carried out under a system reveals the values of the coordinates in an element $dq$ of the configuration space. The function  $\Psi(q)$ is called the wave function.

A knowledge of a  wave function allows in the principle to calculate also  the probabilities of various results of any measurement at all (not only coordinates'measurement), At that, all these probabilities are determined by the 
expressions, which are bilinear on $\Psi(q)$ and $\Psi^*(q)$, the most general form of which is 
\begin{equation}
\label{eq1}
 \int\int\Psi(q)\Psi^*(q') \phi (q,q') dqdq',
\end{equation}
where the function $\phi (q,q')$ depends on the measurement kind and result".
The formulation of the given postulate is given in the quite similar form in all other textbooks, see for instance \cite{Kramers}, \cite{Bohm}, \cite{Davydov}, \cite{Feynman}, \cite{Berklee}. Moreover, just in the given content the  postulate above formulated is used in all modern scientific literature.
At the same time it was drawn the attention in the work
\cite{Optics_Communications} on  the following. Please, the citation from the given work: "Let us also draw attention on that most often used at present
formalism for description of spectroscopic transitions' dynamics in
the frame of the gyroscopic model by means of density operator has a
substantial disadvantage. It really seems to be understandable that in
some experimental cases, in particular, for local single centers, which
can be used in nanotechnology, for instance, for elements of binary
logic, quantum informatics, quantum computing with very fast rates
of processes of information treatment, realized on semiconductor quantum dots, Josephson cubits, spin-qubits and so on, statistical density
operator formalism can be inapplicable for description of transitions'
dynamics at all. The main reason is the following: If the state of
quantum system is a mixed state, that seems to be necessary for
functioning of nanotechnology elements, then the event frequency of
realization of definite pure state in the superposition might not
correspond to its probability. In other words, the well known law of
probabilities theory — the law of large numbers has to be taken into
account. Moreover, by measurements with short times, which are
comparable with the residence time in definite pure state, incoming in
the superposition of the states, the amplitude of given state cannot be
determined by its probability for any quantum system.
New approach for a description of radio- and optical spectroscopic
transitions' dynamics in condensed matter, which allows overcoming
given difficulties, was developed in \cite{Doklady_NANB}".  And further: "Operator equation, describing the optical transitions' dynamics (for the
example of simple 1D-model of quantum system), has been obtained in \cite{Doklady_NANB} by
using of stated transition operators' method. It has been shown, that the given equation is
operator equivalent to Landau–Lifshitz  equation in its difference-differential
form, which takes usual differential form in continuum limit.
In other words, it has been shown, that within the frames of transition
operator method we can obtain the equations for dynamics of
spectroscopic transitions in well approved in the practical studies
form, that is, in the same form like the form of phenomenological Bloch
equations [let us remark, that Bloch
equations are in fact Landau–Lifshitz equations in which the only phenomenological relaxation term was added] or the equations, which were obtained from a density matrix
formalism, [which are also dealing in all modern studies with the notion of a probalility]. In fact, the  transition operator method was unified with a 
gyroscopic model. It was done for the first time to our knowledge and
it indicates on some development of the method at all." It
was demonstrated in the above cited paper, that the transition operator method has a number of
advantages in comparison with other approaches. The main merit of the method is that, that it is dealing with the states of the system immediately but not with the probalities of the presence of the particles in the concrete states. It is especially actual for the description of the processes, for which the notion of the probability itself cannot be used.

 For the work presented, the remark above cited is substantial, since it is followed from here that there are the restrictions on the application of the second main postulate  in the form above presented. Moreover,  it is entirely inapplicable for the description of quantum states  and processes, when [we accentuate once again] the notion of the probability cannot be used. 

Let us concern the history of an introduction of the given postulate in quantum physics. The notion of wave function in quantum physics was introduced by Schr\"{o}dinger \cite{Schroedinger1}, \cite{Schroedinger2},  \cite{Schroedinger3}, \cite{Schroedinger4}, \cite{Schroedinger5}, \cite{Schroedinger6}, \cite{Schroedinger7}. 
Let us remark that Schr\"{o}dinger's theory is a real display of a plurality of the variants of a quantum theory construction, that was strongly mathematically proved in \cite{A_Dovlatova_D_Yerchuck}. Quite other ideas were put in a base by Schr\"{o}dinger for the creation of a new version of quantum mechanics in comparison with Heisenberg matrix quantum mechanics. The  form of quantum theory, developed by Schr\"{o}dinger, has however the phenomenological character, since Schr\"{o}dinger has used the known Hamiltonian analogy between mechanics and optics, which he has extended to include real "physical" or "undulatory" mechanics, instead of mere geometrical mechanics. In its turn, the extension was based on the very interesting and fundamental researches of Louis de Broglie on what he called "phase-waves" ("ondes de phase") and thought to be associated with the motion of material point-like microobjects, especially with the motion of an electron or proton. The Schr\"{o}dinger's viewpoint is that  that   material point-like microobjects consist of, or are nothing but, wave-systems [the given conclusion seems to be correct the only partly, see futher our comment to the given viewpoint]. At the same time, Schr\"{o}dinger does not offer  "the slightest explanation of why a wave-systems' representation of mechanical microobjects of definite mass and charge seem to be realized in the Nature". It is his own comment \cite{Schroedinger7}, that, properly speaking, really indicates on the phenomenological character of the theory proposed. Schr\"{o}dinger writes, that the opposite viewpoint, which neglects altogether the waves discovered by Louis de Broglie and treats only the motion of material point-like microobjects, has led to very big difficulties in the theory of atomic mechanics.
 
We will show, that 
 Schr\"{o}dinger's wave theory is based on the phenomenon of a corpuscular-wave dualism of microworld particles, mapping, mainly, in contrast to Schr\"{o}dinger opinion, its corpuscular aspect. The wave aspect is represented the only in an initial stage. Naturally, it does not explain the physical nature of the given phenomenon, although Schr\"{o}dinger was very near to give the given explanation.

 The absence of the explanation  of the nature of the phenomenon of a corpuscular-wave dualism gave birth in the scientific literature to the some extent mystical opinion that the same elementary particle in quantum microworld, for instance, electron, can be simultaneously corpuscule and wave, and that the given representation  has not any classical analogue in a macroworld. Scientists humbled with the given conclusion, despite on that, that the quantum laws and equations are very similar to the corresponding laws and equations of classical physics. It is followed  from the given resemblance in the  mathematical description, that the resemblance has to be  between classical and quantum objects themselves too. The key to the correct physical explanation of the given seemingly mystical  phenomenon has been found quite recently, see the work \cite{DAA}. The corpuscular-wave dualism has been explained on the example of photon, which like to an any elementary particle has a generic connection with the corresponding field. [It is well known, that according to Standard Model photon is the messenger of electromagnetic field]. We wish to remark, that the explanation of a corpuscular-wave dualism in light phenomena is rather simple. It is connected with the  complex structure of EM-field. Really,  the quantized EM-field represents itself according to the model proposed in \cite{DAA} the discrete massless boson-"atomic" space structure like to an atomic structure of condensed matter.  The origin of waves
in a given structure is determined by the mechanism, quite analogous to the Bloch wave formation in the 
solid state of condensed matter.  At the same time there
are simultaneously the corpuscles, propagating along the given EM-field boson-"atomic" chain
structure. According to the model the light corpuscles represent themselves chargeless spin 1/2 topological relativistic solitons - photons, formed in usual
conditions (or spinless charged solitons in so-called "doped" EM-field structure). In other words, corpuscular and wave properties of light are determined by \textit{different} objects, and it is quite understandable, that the display of the corpuscular or wave nature of light will
be dependent on experimental conditions.  We suggest, that  the given model can be generalised and expanded for the explanation of the   phenomenon of a corpuscular-wave dualism, displayed by the studies of elementary particles with nonzero rest mass values.  The strong support in the favour of the given hypothese is the following. The   Standard
Model, being to be the quantum and relativistic theory which describes in a unified framework the
electromagnetic, weak and strong forces of elementary particles is based on a very powerful principle, local or gauge symmetry: the
fields corresponding to the particles of a given internal symmetry group. In the given model a field is associated to each particle, see, for instance \cite{Djouadi}. In other words, any elementary particle and the field, associated with the given particle represent themselves a single whole, that is a single quantum object. It gives the base for its quantum description within quantum liquid model like to that represented in \cite{DAA} for EM-field description, the corresponding elementary particle in which is photon. Hence, it is followed, that  the field, associated with any elementary particle has an own structure, consisting of separate subobjects including corpuscles and waves. Then the nature of  the   phenomenon of a corpuscular-wave dualism, displayed by the studies of elementary particles with nonzero rest mass values, like to the nature of the same phenomenon for the light, becomes the natural explanation - wave and corpuscular properties are determined by different constituents of associated field-particle quantum object. In other words, the physics of elementary particles seems to be dealing with fields having an own structure, allowing to realize wave and corpuscular phenomena by their different constituents.  

Concerning Schr\"{o}dinger's theory, we can conclude in the light of the premises, that it displays both aspects of quantum phenomena, just corpuscular and to some extent wave aspects, however it has been done independently from each other and phenomenologically. So, there is the field for its development and for the elaboration of a more complete and generalised variant, in which the wave and corpuscular characterisics of rather complicated quantum corpuscular-wave objects will be considered from unified positions, based on a clear representation of the corpuscular-wave dualism  interpreted in \cite{DAA}.
In fact, we have elucidated the general conclusion in \cite {Berklee}, that Schr\"{o}dinger's theory "is not equivalent to the complete quantummechanical theory", being to be to our opinion the very good its first approximation.

Let us concern briefly the principle themselves of the  quantum mechanics construction by Schr\"{o}dinger.  Schr\"{o}dinger took into account the following. The state of any dynamic system in classical mechanics can be described in any moment of time by dimensioning of
coordinate set and conjugate to them impulse components, that is, in the case of $n$
degrees of freedom by $2n$ variables. At the same time for the description  of the state of  a quantum system can be choosed or cordinates or impulses, since cordinates and  impulses cannot be measured simultaneously.

Futher we reproduce almost literally the argumentation of Schr\"{o}dinger, concerning just of an introduction of scalar field function for the description of atomic dynamic systems. 

Schr\"{o}dinger considers at first the simplest example of a mechanical system - a material point, mass $m$ , moving in a conservative field of force $ V(x, y, z)$ . Using the usual notations the kinetic energy $ T$  is

\begin{equation}
\label{eq4}
T = \frac{1}{2}m({\frac{dx}{dt}}^2+{\frac{dy}{dt}}^2+{\frac{dz}{dt}}^2) = \frac{1}{2m}(p^2_x+p^2_y+p^2_z).
\end{equation}
The well-known Hamilton function of action $W$ is
\begin{equation}
\label{eq5a}
W = \int\limits_{t_0}^{t}(T-V)dt,
\end{equation}
which is a function of the upper limit $t$ and of the final values of the coordinates $x, y, z$. It satisfies the Hamilton differential equation in partial derivatives 
\begin{equation}
\label{eq5}\begin{split}
&\partial{W}/\partial{t}+(l/2m)[(\partial{W}/\partial{x})^2+(\partial{W}/\partial{y})^2+(\partial{W}/\partial{z})^2] \\
&+ V(x,y,z) = 0,\end{split}
\end{equation}

To solve the equation (\ref{eq5}), Schr\"{o}dinger represents the action $W$  in the form
\begin{equation}
\label{eq6}
W = -Et + S(x,y,z),
\end{equation}
with $E$ being an integration constant, that is, the total energy, and $S$ a function of $x, y, z$ only.   Equation (\ref{eq5}) may then be written
\begin{equation}
\label{eq6a}
|grad W| =[2m(E-V)]^{\frac{1}{2}},
\end{equation}

Schr\"{o}dinger indicates, that the relation (\ref{eq6a}) has a very simple geometrical interpretation: "Assume $t$ constant for the moment. Any function $W$ of space alone can be described by giving geometrically the system of surfaces on which W is constant and by writing down on each one of these surfaces the constant value, say $W_0$, which the function $W$ takes on it. On the other hand, we can easily construct a solution of Eq.(\ref{eq6a}) starting from an arbitrary surface and an arbitrarily chosen value $W_0$, which we ascribe to it. For after having chosen starting surface and starting value and after-still arbitrarily-having designated one of its two sides or ''shores'' to be the positive value, we simply have to extend the normal at every point of the chosen surface to the length, say
\begin{equation}
\label{eq6ab}
dn = \frac{W_0}{[2m(E-V)]^{\frac{1}{2}}},
\end{equation}
The totality of points arrived at in this way will fill a surface to which we obviously have to ascribe the value $W_0+dW_0$. The continuation of this procedure will supply us the whole system of surfaces and values of constants belonging to them, i.e. the whole distribution in space of the function $W$, at first for $t$ constant.
Now let the time vary, (\ref{eq6ab}) shows that the system of surfaces will not vary, but that the values of the constants will travel along the normals from surface to surface with a certain velocity $u$, given by
\begin{equation}
\label{eq7}
u = \frac{E}{[2m(E-V)]^{\frac{1}{2}}},
\end{equation}
The velocity $u$ is a function of the energy-constant $E$ and besides, since it contains $V(x,y,z)$ is a function of the coordinates.
Instead of thinking of the surfaces being to be fixed in space and letting the values of the constant wander from surface to surface, we may equally well think of a certain numerical value of $W$, attached to a certain individual surface and let the surfaces wander in such a way that each of them continually takes the place and exact form of the following one. Then the quantity $u$ will denote the normal-velocity of any surface at any one of its points. Adopting this view we arrive at a picture which exactly coincides with the propagation of a stationary wave-system in an optically non-homogeneous (but isotropic)
 medium, $W$ being proportional to the phase and $u$ being the phase-velocity. (The index of refraction would have to be taken proportional to $u^{-l}$.) The above-mentioned construction of normals $dn$ is obviously equivalent to Huygens' principle. The orthogonal curves of our system of $W$-surfaces form a system of rays in our optical picture; they are possible orbits of the material point in the mechanical problem. Indeed, it is well known that
\begin{equation}
\label{eq8}
p_x = m\frac{dx}{dt} = \frac{\partial{W}}{\partial{x}},
\end{equation}
(with two analogous equations for $y$ and $z$). It may be useful, to remark, that the phase-velocity $u$ is not the velocity of the material point. The latter is, by (\ref{eq8}) and (\ref{eq5})
\begin{equation}
\label{eq9}
v = \sqrt{(\frac{dx}{dt})^2 + (\frac{dy}{dt})^2 + (\frac{dz}{dt})^2} = (\frac{2(E-V)}{m})^{\frac{1}{2}},
\end{equation}
Comparing (\ref{eq7}) and (\ref{eq9}) we see, that $u$ and $v$ vary even inversely to each other. The well-known mechanical principle due to and named after Hamilton can very easily be shown to correspond to the equally well-known  optical  principle  of  Fermat".

Schr\"{o}dinger remarks further: "Nothing of what has hitherto been said is in any way new. All this was very much better known to Hamilton himself than it is in our day to a good many physicists. Indeed, the theory of the propagation of light in a non-homogeneous medium, which Hamilton had developed about ten years earlier, became, by the striking analogy which occurred to him, the starting-point for his famous theories in pure mechanics. Notwithstanding the great popularity reached by the latter, the way which had led to them was nearly forgotten". Schr\"{o}dinger draws attention that "though in  above-stated reasoning the conceptions "wave-surfaces," "Huygens principle," "Fermat's principle" come into play, nevertheless the whole established analogy deals rather with geometrical optics than with real physical or undulatory optics. Indeed the chief and fundamental mechanical conception is that of the path or orbit of the material particle, and it corresponds to the conception of rays in the optical analogy. Now the conception of rays is thoroughly well-defined only in pure abstract geometrical optics". The situation is quite different, when "the dimensions of the beam or of material obstacles in its path become comparable with the wavelength.    And even when this is not the case, the notion of rays is, in
physical optics, merely an approximate one. It is wholly incapable of being applied to the fine structure of real optical phenomena, for example, to the phenomena of diffraction. Even in extending geometrical optics somewhat by adding the notion of Huygens' principle in the simple form the most simple phenomena of diffraction cannot be accounted for without adding some further very strange rules concerning the circumstances under which Huygens' envelope-surface is or is not physically significant". Schr\"{o}dinger bears in mind the construction of "Fresnel's zones" and he argues further: "These rules would be wholly incomprehensible to one versed in geometrical optics alone. Furthermore, it may be observed that the notions which are fundamental to real physical optics, that is the wave-function itself ($W$ is merely the phase), the equation of wave-propagation, the wavelength and frequency of the waves, do not enter at all into the above stated analogy. The phase-velocity $u$ does enter but we have seen that it is not very intimately connected with the mechanical velocity $v$.
At first sight it does not seem at all tempting, to work out in detail the Hamiltonian analogy like to in real undulatory optics. By giving the wave-length a proper well-defined meaning, the well-defined meaning of rays is lost at least in some cases, and by this the analogy would seem to be weakened or even to be wholly destroyed for those cases in which the dimensions of the mechanical orbits or their radii of curvature become comparable with the wave-length. To save the analogy it would seem necessary to attribute an exceedingly small value to the wavelength, small in comparison with all dimensions that may ever become of any interest in the mechanical problem. But then again the working out of an undulatory picture would seem superfluous, for geometrical optics, is the real limiting case of undulatory optics for vanishing wavelength". In fact, the given words is the attempt of Schr\"{o}dinger to understand the nature of a corpuscular-wave dualism of the objects of the microworld. It is seen, that his insight in the physical nature of corpuscular-wave dualism  is rather vague, that has left an imprint on all the theory. Nevertheless our analysis of Schr\"{o}dinger's theory (see, further for details) shows, that Schr\"{o}dinger was very near to the concept of quantum liquid structure of EM-field, allowing to understand the nature of a corpuscular-wave dualism and to confirm his interpretation of a field scalar function.

Further, Schr\"{o}dinger formulates target setting of the construction  of quantum mechanics and the way of its phenomenological solution based on the comparison with wave optical phenomena: "Now compare with these considerations the very striking fact, of which we have today irrefutable knowledge, that ordinary mechanics is really not applicable to mechanical systems of very small, viz. of atomic dimensions. Taking into account this fact, which impresses its stamp upon all modern physical reasoning, is one not greatly tempted to investigate whether the non-applicability of ordinary mechanics to micro-mechanical problems is perhaps of exactly the same kind that the non-applicability of geometrical optics to the phenomena of diffraction or interference and may, perhaps, be overcome in an exactly similar way? The conception is: the Hamilton analogy has really to be worked out towards undulatory optics and a definite size is to be attributed to the wave-length in every special case. This quantity has a real meaning for the mechanical problem, that is, ordinary mechanics with its conception of a moving point and its linear path (or more generally of an "image-point" moving in the coordinate space) is only approximately applicable in the case of a path,  whose radii of curvature are large in comparison with the wave-length". 

We have to remark, that optical phenomena of diffraction and interferention are rather subtle, its appearance is dependent on the light intensity level. It has to be taken into account by the development of Schr\"{o}dinger viewpoint, see further.

  Schr\"{o}dinger continues: "If this is not the case, it is a phenomenon of wave-propagation that has to be studied. In the simple case of one material point moving in an external field of force the wave-phenomenon may be thought to be taking place in the ordinary three-dimensional space; in the case of a more general mechanical system it will primarily be located in the coordinate space (q-space, not pq-space) and will have to be projected somehow into ordinary space. At any rate the equations of ordinary mechanics will be of no more use for the study of these micro-mechanical wave-phenomena than the rules of geometrical optics are for the study of diffraction phenomena. Well known methods of wave-theory, somewhat generalized, lend themselves readily. The conceptions, roughly sketched in the preceding are fully justified by the success which has attended their development".

 Further, it is followed the  pioneering Schr\"{o}dinger's idea of quantum \textit{wave} mechanics formulation. "Let us return to the system of $W$-surfaces",-writes Schr\"{o}dinger,- "and let us associate with them the idea of stationary sinusoidal waves whose phase is given by the quantity $W$, Eq.(\ref{eq5}). The wave-function, say $\psi(x,y,z,t)$ will be of the form
\begin{equation}
\label{eq10}\begin{split}
&\psi(x,y,z,t) = A(x,y,z)\sin(W/K)= \\
&A(x,y,z)\sin[\frac{-Et}{K}+\frac{S(x,y,z)}{K}],
\end{split}\end{equation} 
where $A(x,y,z)$ is an "amplitude" function. The constant $K$ must be introduced and must have the physical dimension of action (energy$\times$time), since the argument of a sine must always be a pure number". "Now", - Schr\"{o}dinger argues, - "since the frequency of the wave (\ref{eq10}) is obviously
\begin{equation}
\label{eq11}
\nu = \frac{E}{2\pi K},
\end{equation}
 supposing $K$ to be \textit{a universal constant}, independent on $E$ and independent on the nature of the mechanical system, because if this be done and $K$ be given by the value $h/2W$, then the frequency $\nu$ will be given by
\begin{equation}
\label{eq12}
\nu = \frac{E}{h},
\end{equation}
where $h$ is Planck constant. Thus, the well known universal relation between energy and frequency is arrived at in a rather simple and unforced way.

In ordinary mechanics the absolute value of the energy has no definite meaning, only energy-differences have. This difficulty can be met and a zero-level of energy can be defined in an entirely satisfactory way by using relativistic mechanics and the conception of equivalence of mass and energy. But it is unnecessary to dwell on this subject here. While the frequency $\nu$  is indeed dependent on the zero-level of energy, their wave-length is not. And after what has been said above, it is the wave-length that is of greatest interest. The comparison of this quantity with the dimensions of the path or orbit of the material particle, calculated according to ordinary mechanics, will tell us whether the latter calculation is or is not of physical significance, whether the methods of ordinary mechanics are approximately applicable to the special problem or not.   The wave-length $\lambda$ according to (\ref{eq12}) and (\ref{eq7}) is
\begin{equation}
\label{eq13}
\lambda = \frac{u}{\nu} = \frac{h}{[2m(E-V)]^{\frac{1}{2}}},
\end{equation}
where $E-V$ is the kinetic energy, which indeed is independent of the zero-level of the total energy.  Inserting its value we have
\begin{equation}
\label{eq14}
\lambda = \frac{u}{\nu} = \frac{h}{mv}.
\end{equation}
To test the question whether an electron, moving in a Keplerian orbit of atomic dimensions may, following our hypotheses, still be dealt with by ordinary mechanics, let a be a length of atomic dimensions and compare $\lambda$ with $a$
\begin{equation}
\label{eq15}
\frac{\lambda}{a} = \frac{h}{mva}.
\end{equation}
The denominator on the right is certainly of the order of magnitude of the moment of momentum of the electron, and the latter is well known to be of the order of magnitude of Planck's constant for a Keplerian orbit of atomic dimensions. So $\frac{\lambda}{a}$ becomes of the order of unity and, following our conceptions, ordinary mechanics will be no more applicable to such an orbit than geometrical optics is to the diffraction of light by a disk of diameter equal to the wave-length. Were a physicist to try to understand the latter phenomenon by the conception of rays, with which he is acquainted from macroscopic geometrical optics, he would meet with most serious difficulties and apparent contradictions. The "rays" (stream lines of the flow of energy) would no longer be rectilinear and would influence one another in a most curious way, in full contradiction with the most fundamental laws of geometrical optics. In the same way the conception of orbits of material points seems to be inapplicable to orbits of atomic dimensions. It is very satisfactory, that the limit of applicability of ordinary mechanics is, by equating $K$ (essentially)
to Planck's constant, determined to an order of magnitude, which is exactly the one to be postulated, if the new conception is to help us in our quantum difficulties. We may add, that by Eq.(\ref{eq12}) for a Keplerian electronic orbit of the order of magnitude of a high quantum orbit, the relation of wave-length to orbital dimensions becomes of the order of magnitude of the reciprocal of the quantum number. Hence ordinary mechanics will offer a better and better approximation in the limit of increasing quantum number (or orbital dimensions), and this is just what is to be expected from any reasonable theory".

 Here we have to remark, that just Eq.(\ref{eq10}) is the starting qenuine wave equation, which allows to describe the wave properties of rather complicated systems of quantum fields, which, being to be represented by quantum liquids allow to represent the model of atoms being to be the superposition of the corpuscles of corresponding fields in spirit of Standard Model, that is the superposition of the field system, corpuscles in which are elementary particles producing the atomic nuclei - protons and neutrons, for which the strong interactions are responsible and the field system, corpuscles in which are electrons, for which the electromagnetic interactions are responsible. In fact, Schr\"{o}dinger has put into consideration two different field scalar functions, the first of which is determined by  Eq.(\ref{eq10}).

The following arguments gives Schr\"{o}dinger to put into consideration the second field scalar function and to obtain the equation for its determination: "By the fundamental equation  Eq.(\ref{eq10}) the phase velocity $u$ given by Eq.(\ref{eq7}) proves to be dependent on the frequency $\nu$. Therefore Eq.(\ref{eq7}) is an equation of dispersion. By this a very interesting light is thrown on the relation of the two velocities - velocity $v$ of the moving particle and   phase-velocity $u$,  $v$ is easily proved to be exactly the so-called group velocity belonging to the dispersion formula Eq.(\ref{eq7}). By using this interesting result it is possible to form an idea how ordinary mechanics is capable of giving an approximate description of our wave motion. By superposing waves of frequencies in a small interval ($v$, $v+dv$) it is possible to construct a "parcel of waves" the dimensions of which are in all directions rather small, though they must be rather large in comparison to the wave-length. Now it can be proved, that the motion of - let us say - the "center of gravity" of such a parcel will, by the laws of wave propagation, follow exactly the same orbit which the material point would have by the laws of ordinary mechanics. This equivalence is always maintained, even if the dimensions of the orbit are not large in comparison with the wave-length. But in the latter case it will have no significance, the wave parcel being spread out in all directions far over the range of the orbit. On the contrary, if the dimensions of the orbit are comparatively large, the motion of the wave parcel, considered to be  a whole may afford a sufficient idea of what really happens, if we are not interested in its intrinsic constitution". Schr\"{o}dinger continues: "This "motion as a whole" is governed by the laws of ordinary mechanics, the wave-phenomena must in this case be studied in detail. This can only be done by using an "equation of wave propagation". Which one is this to be?". 

In fact, on the one hand, it is an attempt by Schr\"{o}dinger  himself to explain   the corpuscular-wave dualism, that is, the attempt to deduce the appearance of the corpuscles from a wave field. It is quite phenomenological, although it is to some extent succesfull, since it gives the correct value of the velocity of the given quasiparticle. The given  little success seems to be not ocassional, since the real particle, for example, photon, which, being to be topological soliton, can be modelled analogously, the only with appropriate choosing of an envelope function. On the other hand, it is the direct indication that the subsequent consideration of the task of the description of the dynamics of atomic systems will display the corpuscular aspect. In fact Schr\"{o}dinger himself understands (maybe on semiintuitive level), that his theory is in fact the field theory, we have to add, that the only on its starting stage. The proof for the given viewpoint is the following.  Please, Schr\"{o}dinger's own comment to the name both for differential equation for  function $|\Psi(q_1,q_2,...,q_N, t)|^2$ and for the function $\Psi(q_1,q_2,...,q_N, t)$ itself. In the fourth (last) part of the series of works "Quantisierung als Eigenwertproblem" Schr\"{o}dinger writes \cite{Schroedinger4}: "Wenn wir also Gleichung (1) oder (1') [that is, in the modern terminology for both  the stationary  Schr\"{o}dinger equation, given by  (\ref{eq18}) (see further), and the nonstationary  Schr\"{o}dinger equation, the dependence of  $\Psi(q_1,q_2,...,q_N, t)$ on a time in which is determined by (\ref{eq25}) (see further), we wish to remark, that the only given equations are used in the  modern nonrelativistic quantum mechanics in the linear case] gelegentlich als Wellengleichung bezeichnet haben, so geschah das eigentlich zu Unrecht, sie w\"{a}re richtiger als "Schwingungs-" oder "Amplituden"-gleichung zu bezeichnen. Wir fanden aber mit ihr das Auslangen, weil ja an diese das Sturm-Liouvillesche Eigenwertproblem sich kn\"{u}pft - ganz ebenso wie bei dem mathematisch v\"{o}llig analogen Problem der freien Scbwingungen von Saiten und Membranen - und nicbt an die eigentliche Wellengleichung". In the cited work Schr\"{o}dinger introduces the new name  for the function $\Psi(q_1,q_2,...,q_N, t)$ - "Feldskalar" [field scalar] instead of "wave function", just to the given name Schr\"{o}dinger gives preference. It indicates to our opinion on rather deep his semiintuitive insight  in the field, indicating on its understanding of the relation of the theory developed with the field theory and on its corpuscular aspect description mainly. 

Further we will represent very briefly the argumentation of Schr\"{o}dinger by obtaining his famous equations.

Schr\"{o}dinger writes \cite{Schroedinger2}, \cite{Schroedinger4}, \cite{Schroedinger7}:
"Suppose, the extreme values of the following integral extending over all space were required.
\begin{equation}
\label{eq16}
\begin{split}
&I_1 = \int\int\int\frac{h^2[(\frac{\partial\Psi(x,y,z)}{\partial x})^2 
+ (\frac{\partial\Psi(x,y,z)}{\partial y})^2]}{8\pi^2 m} dxdydz +\\
&\int\int\int\{\frac{h^2(\frac{\partial\Psi(x,y,z)}{\partial z})^2}{8\pi^2 m} + V(x,y,z)\Psi^2(x,y,z)\}  dxdydz,
\end{split}
\end{equation}
 all single-valued, finite and continuously differentiable functions $\Psi$ being admitted to concurrence" that give the following "normalizing" integral a constant value, say 1:
\begin{equation}
\label{eq17}
 I_2 = \int\int\int \Psi^2(x,y,z)  dxdydz = 1.
\end{equation}

In carrying out the variation under this "accessory condition" in the well known manner, the following  equation is found being to be the well known necessary condition for an extreme value of integral (\ref{eq16})

\begin{equation}
\label{eq18}
 \triangle\Psi(x,y,z) + \frac{8\pi^2[E - V(x,y,z)]}{h^2} = 0,
\end{equation}
 the constant $- E$ being the Lagrangian multiplier with which the variation of the second integral has to be multiplied and added to the first, up to take care of the accessory condition. Thus the normalized characteristic functions of (\ref{eq18}) are exactly the so-called extremals of the integral (\ref{eq16}) under the normalizing condition (\ref{eq17}), whereas the characteristic values, that is the values, that are admissible for the constant $E$ are nothing else than the extreme values of integral (\ref{eq16}). 
Now the integrand of (\ref{eq16}) proves on closer inspection to have a very simple relation to the ordinary Hamiltonian function of our mechanical problem - in the sense of ordinary mechanics.    The said function is:
\begin{equation}
\label{eq19}
 \frac{1}{2m}(p^2_x + p^2_y + p^2_z) + V(x,y,z).
\end{equation}

Take this function to be a homogeneous quadratic function of the momenta $p_x, p_y, p_z$ and of unity and replace therein $p_x, p_y, p_z, 1$ by $\frac{h}{2\pi}(\frac {\partial\Psi}{\partial x})$, $\frac{h}{2\pi}(\frac {\partial\Psi}{\partial y})$, $\frac{h}{2\pi}(\frac {\partial\Psi}{\partial z})$, $\Psi(x,y,z)$ respectively. There results the integrand of (\ref{eq16}). This immediately suggests extending our variation problem and hereby our wave-equation (\ref{eq18}) to a wholly arbitrary conservative mechanical system. The Hamiltonian function will be of the form
\begin{equation}
\label{eq20}
\frac{1}{2}\sum_{l=1}^{N}\sum_{k=1}^{N} a_{lk}(q_1,q_2,...,q_N) p_l p_k + V(q_1,q_2,...,q_N),
\end{equation}
with $a_{lk}(q_1,q_2,...,q_N) = a_{kl}(q_1,q_2,...,q_N)$. Take (\ref{eq20}) to be a homogeneous quadratic function of $p_1, p_2, ..., p_N, 1$ and replace these quantities by  $\frac{h}{2\pi}(\frac {\partial\Psi}{\partial q_1})$, ... , $\frac{h}{2\pi}(\frac {\partial\Psi}{\partial q_N})$, $\Psi(q_1,q_2,...,q_N)$ respectively.   Writing $\triangle_p$ for the determinant
\begin{equation}
\label{eq21}
\triangle_p = \mid \sum\pm a_{lk}(q_1,q_2,...,q_N \mid
\end{equation}
we form the integral
\begin{equation}
\label{eq22}\begin{split}
&I_1 = \int ... \int [\frac{h^2}{8\pi^2}\sum_{l=1}^{N}\sum_{k=1}^{N} a_{lk}(q_1,q_2,...,q_N)\times\\
&\frac {\partial\Psi(q_1,q_2,...,q_N)}{\partial q_l}\frac {\partial\Psi(q_1,q_2,...,q_N)}{\partial q_k}) +\\ &V(q_1,q_2,...,q_N)\Psi^2(q_1,q_2,...,q_N)] \triangle_p^{-\frac{1}{2}}dq_1,dq_2,...,dq_N ,\end{split}
\end{equation}
taken over the whole space of coordinates and seek its extreme values under the accessory condition
\begin{equation}
\label{eq23}
 I_2 = \int ... \int \Psi^2(q_1,q_2,...,q_N) \triangle_p^{-\frac{1}{2}}dq_1,dq_2,...,dq_N = 1 .
\end{equation}
This leads to the generalization of Eq.(\ref{eq18}), viz.
\begin{equation}
\label{eq24}\begin{split}
&\triangle_p^{\frac{1}{2}}\sum_{l=1}^{N}\frac{\partial}{\partial q_l}[\triangle_p^{-\frac{1}{2}}\sum_{k=1}{N}a_{lk}(q_1,q_2,...,q_N)\frac{\partial\Psi(q_1,q_2,...,q_N)}{\partial q_k}\\
&+ \frac{8\pi^2[E - V(q_1,q_2,...,q_N)]}{h^2}\Psi(q_1,q_2,...,q_N) = 0,
\end{split}
\end{equation}
$- E$ being the Lagrangian multiplier. The double sum appearing in Eq.(\ref{eq24}) is a sort of generalized Laplacian in the N-dimensional, non-euclidean space of coordinates. The necessary appearance of $\triangle_p^{-\frac{1}{2}}$ in an integrals (\ref{eq22}) and (\ref{eq23}) is well known from Gibbs' statistical mechanics; $\triangle_p^{-\frac{1}{2}}dq_1,dq_2,...,dq_N$ is simply the non-euclidean element of volume, for example $r^2 \sin\theta d\theta d\phi dr$ in the case of one material point of unit mass, whose position is fixed by three polar coordinates $ r$ , $ \theta$, $ \phi$. (In omitting the determinant the integrals would not be invariant relative to point transformations; they would depend on the choice of generalized coordinates.)" 

Now we will represent the discussion on 
 the real physical meaning of the wave-function $\Psi(q_1,q_2,...,q_N)$. Schr\"{o}dinger remarks, that
 "Eq.(\ref{eq18}) or in the more general case, Eq.(\ref{eq24}) give the dependence of the wave-function $\Psi(q_1,q_2,...,q_N)$ on the coordinates only, the dependence on time being given for every one particular solution, corresponding to a particular characteristic value $E = E_l$ by the real part of
\begin{equation}
\label{eq25}
 \exp[(\frac{2\pi E_lt}{h} + \theta_l)i],
\end{equation}
the $\theta_l$ being phase constants. So if $\Psi_n(q_1,q_2,...,q_N)$, $n \in N$,  be the characteristic functions the most general solution of the wave-problem will be (the real part of)
\begin{equation}
\label{eq26}\begin{split}
&\Psi_n(q_1,q_2,...,q_N, t)= \\
&\sum_{n=1}^{\infty} c_n\Psi_n(q_1,q_2,...,q_N) \exp[(\frac{2\pi E_lt}{h} + \theta_l)i].\end{split}
\end{equation}
(For simplicity's sake the characteristic values are supposed  to be all single and discrete.) The set $\{c_n\}, n \in N$  are real constants. Now form the square of the absolute value of the complex function $\Psi_n(q_1,q_2,...,q_N, t)$. 
\begin{equation}
\label{eq27}\begin{split}
&\Psi_n(q_1,q_2,...,q_N, t)\Psi^*_n(q_1,q_2,...,q_N, t) = \\
&|\Psi_n(q_1,q_2,...,q_N, t)|^2 = 2\sum_{n=1}^{\infty} \sum_{m=1}^{\infty}c_nc_m\Psi_n(q_1,q_2,...,q_N)\times\\ &\Psi^*_m(q_1,q_2,...,q_N)\cos[\frac{2\pi(E_n - E_m)t}{h} + \theta_n - \theta_m].
\end{split}
\end{equation}

This of course, like $\Psi_n(q_1,q_2,...,q_N, t)$ itself, is in the general case a function of the generalized coordinates $q_1, q_2 ,..., q_N$ and the time $t$, - not a function of ordinary space and time. This raises some difficulty in attaching a physical meaning to the wave-function. In the case of the hydrogen atom (a one-body problem) the difficulty disappears. In this case it has been possible to compute fairly correct values for the intensities, for example, of the Stark effect components  by the following hypothesis: the charge of the electron is not concentrated in a point, but is spread out through the whole space, proportional to the quantity $|\Psi_n(q_1,q_2,...,q_N, t)|^2$.

It has to be born in mind, that by this hypothesis the charge is nevertheless restricted to a domain of, say, a few Angstroms, the wave-function $\Psi_n(q_1,q_2,...,q_N, t)$ practically vanishing at greater distance from the nucleus. The fluctuation of the charge will be governed by Eq.(\ref{eq27}), applied to the special case of the hydrogen atom".

Let us comment the hypothesis  on the charge distribution, described by continuous function. On the one hand, it is direct consequence of the field character of Schr\"{o}dinger  theory. On the second hand, Schr\"{o}dinger in fact suggests, that the fields associated with elementary particles incoming in the atom structure have 
a new observable quantity - continuosly distributed in a space  the scalar charge function. The given hypothesis was recently proved in the works \cite{Yearchuck_Alexandrov_Dovlatova}, \cite{A_Dovlatova_D_Yerchuck} for the case of an electromagnetic field. Consequently the conclusion on charge spread out through the whole space, proportional to the quantity $|\Psi_n(q_1,q_2,...,q_N, t)|^2$ seems to be correct up to distances, being to be comparable with a  nuclei size, since by a small distances for the charge distribution the strong interaction forces are responsible. On the third side, the indication, that the charge distribution is  restricted to a domain of a few Angstroms, and that  the  scalar function $\Psi_n(q_1,q_2,...,q_N, t)$ practically vanishing at greater distance from the nucleus confirms additionally, that the given part of Schr\"{o}dinger theory is really describes the corpuscular aspect in the dual picture considered.

Further Schr\"{o}dinger gives the generalization of the concept above considered: "Now how are these conceptions to be generalized to the case of more than one, say of N, electrons? Here Heisenberg's formal theory has proved most valuable. It tells us though less by physical reasoning than by its compact formal structure that equation  giving a rectangular component of total electric moment has to be maintained with the only differences that (1) the integrals are 3N-fold instead of three fold, extending over the whole coordinate space; (2) $e z$ has to be replaced by the sum $\sum e_i z_i$ i.e. by the $z$-component of the total electrical moment which the point-charge model would have in the configuration $(x_1, y_1, z_1; x_2, y_2, z_2; ...;  x_N, y_N, z_N)$ that relates to the element $dx_1, ...; dz_N$ of the integration". It was taken into account the intimate connection proved by Schr\"{o}dinger between the matrix and his own theories.  "The achievement of the present theory - which may be imperfect  in many respects" - writes Schr\"{o}dinger - "seems to me to be that by a definite localization of the charge in space and time we are able from ordinary electrodynamics really to derive both the frequencies and the intensities and polarizations of the emitted light. All so-called selection principles automatically result from the vanishing of the triple integral for  the electric dipole moment in the particular case".

The argumentation aforegiven allowed to Schr\"{o}dinger to formulate the following hypothesis concerning the physical meaning of the field skalar $\Psi(q_1,q_2,...,q_N, t)$ in the case of N-electron system: "The real continuous partition of the charge is a sort of mean of the continuous multitude of all possible configurations
of the corresponding point-charge model, the mean being taken with the quantity  $|\Psi_n(q_1,q_2,...,q_N, t)|^2$ representing itself a sort of weight-function in the configuration space.
No very definite experimental results can be brought forward at present in favour of this generalized hypothesis. But some very general theoretical results on the quantity  $|\Psi_n(q_1,q_2,...,q_N, t)|^2$ persuade me that the hypothesis is right. For example, the value of the integral of  $|\Psi_n(q_1,q_2,...,q_N, t)|^2$, taken over the whole coordinate space proves absolutely constant (it should be, if  is a reasonable weight function) not only with a conservative, but also with a non-conservative system".

Therefore, it is clear, that the Schr\"{o}dinger's interpretation of the field skalar $\Psi(q_1,q_2,...,q_N, t)$, introduced in quantum theory by himself, differs 
drastically from its interpretation in modern textbooks on quantum mechanics above
worded. Schr\"{o}dinger does not connect $|\Psi(q_1,q_2,...,q_N, t)|^2$ with a probability at all. We accentuate once again, that the notion  of the probability cannot be used for fast passing  physical processes. At the same time, it can be argued, that the Schr\"{o}dinger's interpretation can be retained even in the given case. 

 According to the proposal of Born \cite{Born}, the field skalar $\Psi(q_1,q_2,...,q_N, t)$
was used to describe the amplitude of the probability of finding the electron in space.
The statistical interpretation of quantum theory, proposed by Born, rejects describing a single event
but only the probabilities in repeated experiments. This is “too big of a sacrifice” according to
Schr\"{o}dinger opinion, and is responsible for some of the major problems of the foundations of the whole theory.

Thereupon, it seems to be interesting also Dirac comments to the Schr\"{o}dinger's theory. Dirac accentuates \cite{Dirac_1}, that the differential equation for the function $\Psi$   
is very closely connected with the equation of Hamilton-Jacobi. Just, if the equation
\begin{equation}
\label{eq28}\begin{split}
\mathcal{H}(q_r, p_r) - W = 0
\end{split}
\end{equation}
is the equation of Hamilton-Jacobi of a system, at that $q_r$, $p_r$  are the canonical variables, then the differential equation for the function $\Psi$ is
\begin{equation}
\label{eq29}\begin{split}
\{\mathcal{H}(q_r, i\hbar\frac{\partial}{\partial q_r}) - W \}\Psi = 0
\end{split}
\end{equation} 
In other words, writes Dirac - "each impulse $p_r$ in the equation of Hamilton-Jacobi is replaced by an operator $i\hbar\frac{\partial}{\partial q_r}$ and 
Schr\"{o}dinger considers the values of parameter $W$, for which exists the comtinuous, single-valued, bounded function $\Psi$ in all $q$-space being to be the energy levels of the system". "He shows" - writes further Dirac -"that if the general solution of the differential equation (\ref{eq29}) is known, then it is easily to find matrices, representing $p_r$ and $q_r$, which are satisfying to all the conditions, which are required by Heisenberg matrix mechanics, therefore the mathematical equivalence of both the theories was proved". Let us remark that  the time is considered in the Schr\"{o}dinger's theory, being to be c-number, instead of, rather than to consider it symmetrically with  the space coordinates. In given meaning  the Schr\"{o}dinger's theory carries the semiquantum character, see analogous remark of Dirac \cite{P.Dirac} concerning his own EM-field quantisation theory and corresponding remark of Schr\"{o}dinger himself \cite{Schrodinger}:
 "Ich m\"{o}chte wiederholen, dass wir eine QM [Quantum Mechanik],
deren Aussagen nicht f\"{u}ur scharf bestimmte Zeitpunkte
gelten soIlen, nicht besitzen. Mir scheint,
dass dieser Mangel sich gerade in jenen Antinomien
kundgibt. Womit ich nicht sagen will, dass es der
einzige Mangel ist, der sich in ihnen kundgibt. Dass die "scharfe Zeit" eine Inkonsequenz innerhalb der QM ist und dass ausserdem, sozusagen
unabh\"angig davon, die Sonderstellung der
Zeit ein schweres Hindernis bildet f\"{u}r die Anpassung
der QM an das Relativit\"{a}tsprinzip, darauf
habe ich in den letzten Jahren immer wieder hingewiesen \cite{SchrodingerA} \cite{SchrodingerB}, \cite{SchrodingerC},
1eider ohne den Schatten eines branchbaren
Gegenvorschlags machen zu k\"{o}nnen" [I would wish to repeat, that we don't have QM, the  substance of which would be regarded to not strictly determined time moments. It seems to me, that the given disadvantage (demerit) is displayed in its contradictions. I don't wish to say, that the given demerit is only one, which is revealed in them. I have time and again pointed out in the last years, \cite{SchrodingerA} \cite{SchrodingerB}, \cite{SchrodingerC}, unfortunately do not making the least counter-offer, that the  "exactly-defined (sharp) time" is the inconsequence inside of QM, and that moreover, so to speak, independently because of that, the special status of the time leads to an impediment in matching of QM with the relativity principle].

It seems to be interesting, that concerning the derivation of Schr\"{o}dinger's equation itself the situation is quite another. Dirac reconsidered in \cite{Dirac_1} the given result of Schr\"{o}dinger's theory from fully quantum positions. Taking into account the resemblance of Schr\"{o}dinger differential equation for the function $\Psi$   
 with the  classsical equation of Hamilton-Jacobi he has obtained Schr\"{o}dinger differential equation in the case of including to canonical variables the time $t$ and corresponding to time canonically conjugated variable $-W$.
It is proof, that Schr\"{o}dinger differential equation for the function $\Psi$ has pure quantum character (but not semiquantum).

Dirac gives the comment to aforeindicated Born suggestion, conserning  the statistical interpretation of Schr\"{o}dinger's quantum theory in \cite{Dirac_2}, he writes directly that the notion of the probability by no means enters in a definitive description of the mechanical processes. The probabilistic description of the mechanical processes is possible, according to Dirac opinion, the only in the case, if the initial information is given  already on  a probability language.

By the way just in the given work Dirac  indicates, that the transitions of the quantum system from one state to another, in particular, from an equilibrium state to an excited state are not instantaneous and that part of the time, related to a duration of stay between the states can be determined theoretically. The given conclusion is concerned of the development of the quantum theory in the direction of the spectroscopic transitions with finite time of transition processes themselves, taking into account of which has allowed recently to explain rather unusual spectroscopic resonance absorption characteristics in carbon nanotubes and in superconducting ceramics \cite{PC}.

Unclear understanding of the nature of corpuscular-wave dualism of quantum systems, the explanation  of which was put in mathematically correct form in the theory of Schr\"{o}dinger [however, we have to mention once again, that Schr\"{o}dinger himself had the rather vague representation on the origin of the given phenomenon and he has described it phenomenologically] leads in modern literature to a mystical treatise of experimental results. Let us give the example of a given treatise. We will cite the comment from review article \cite{Misochko} to rather interesting results referred to quantum field physics: "An international research group led by Ian Walmsley at Oxford was the first to provide experimental evidence that a lattice can be in a state violating the Bell inequalities. In the experiment, phonons excited in the process of spontaneous Raman scattering in two spatially separated diamonds were brought to an entangled state \cite{Lee_1}. The optical phonons that exhibited entanglement were of the symmetry  and had a frequency of 40 THz. To prove entanglement, one of its quantitative characteristics, concurrence, was used. The entanglement of  phonons followed from the positive sign of concurrence; the value of this characteristic,  was established with a reliability above 98 percents. Because phonon counting detectors are currently nonexistent, the authors of \cite{Lee_1} assumed that phonon creation/annihilation events are local and inferred the entanglement of lattice states from the entanglement of Stokes and anti-Stokes photons, whose interference pattern visibility was close to that of maximally entangled Bell states.

The crystals used in the Oxford experiment ('upper' and 'lower') were  in size and  apart; the experiment was conducted at room temperature. The high energy of the phonons studied  ensured that the lattice of both diamonds was in the ground (vacuum) state prior to excitation. The pump laser pulses  passed through a symmetric beamsplitter and arrived at the two crystals via different optical paths, and then, after passing through the crystals, merged into a single pulse, thereby 'erasing' information about the path the photon took. Each of the pump photons could arrive at either the upper or the lower crystal after leaving the beamsplitter". Futher the reviewer elucidates the position of the authors of \cite{Lee_1}: "According to the laws of quantum mechanics, it is impossible to predict before the measurement which way the photon takes, the reason being that it is in a superposition of its two possible states. If photons obeyed classical laws, then, after leaving the beamsplitter, they would move up or down, but by no means in both directions simultaneously. When, upon leaving the beamsplitter, a photon enters the diamond, part of its energy can be absorbed to produce a phonon in the crystal lattice. Because phonons also  quasiparticles, it follows that the two diamonds that have absorbed a photon coming from the beamsplitter share this single phonon, and are therefore entangled. According to classical thinking, the phonon is in the upper or the lower diamond, whereas according to quantum mechanics, it is 'smeared out' over the two". 

Therefore, the representation of the authors of \cite{Lee_1} becomes to be quite clear - one phonon being to be quasiparticle  of rather small size belongs simultaneously to two 
crystals which are distanced from each other in 15 cm (!). It contradicts to  common sense and gives rise to the mysticism. Really let us compare the size of phonon with intercrystal distance. The authors of \cite{Lee_1} evaluate the phonon size, being to be inclusive of $\sim 10^{16}$ atoms, please, the quotation: "The optical phonons are macroscopic,
persistent excitations distributed over
$\sim 10^{16}$ atoms within the crystals". The given evaluation, being to be strongly
overvaluation [on 14-15 orders of value; see further] indicates nevertheless, that the size of phonon is very small in comparison with diamond crystal size, all the more,
in comparison with  the intercrystal distance. Taking into account the foregoing argumentation on the resemblance of classical and quantum laws, equations, objects and phenomena themselves, we can insist, that never can  a single phonon belong to two separare crystals. Let us now continue the comment of the author of \cite{Misochko} and then give the correct qualitative explanation of these very interesting experimental results. 

The author of \cite{Misochko} writes: "In the process of stimulated Raman scattering, the absorbed photon is re-emitted at a lower frequency, and this 'reddened' Stokes photon signals the achievement of an entangled state. However, because the output response of a single photon detector does not identify the exact crystal through which the photon passed, the excited phonons in one of the crystals are quantum-correlated with those in the second. In other words, once the photon is absorbed, the atoms of both crystals are set into motion. To see this, a probe pulse polarized orthogonally to the pump pulse was used, which again was sent with a delay of  through a symmetric beamsplitter onto two diamonds at once. When encountering a phonon, such a probe photon increases in energy (transforms into a 'blue' anti-Stokes photon) and is directed to two detectors via a system of beamsplitters (quarter- and half-wavelength plates). The fact that it is not known in which diamond the phonon is located allows distinguishing the classical state of the two crystals from their quantum (entangled) state. Classically, after sending the probe pulse, the appearance of the blue anti-Stokes photon at the output of the system should be registered with the same probability by both detectors; but if an object is described by the laws of the quantum world, the photon should be registered by one and only one designated detector, because the appearance of a blue anti-Stokes photon should correlate with the appearance of a red Stokes phonon.
It is precisely this nonclassical correlation that the Oxford physicists observed. When a phonon is annihilated and hence atomic motion ceases in one of the crystals, the atoms in the other also suddenly cease moving, although the crystals are spatially apart and cannot interact. And although such entanglement creation and registration experiments lasted a mere  (phonons in diamond are short lived at room temperature), this suffices, if not to store quantum data, then at least to perform quantum computations. Thus, the existence of quantum entanglement is confirmed by measuring correlations between the polarization states of Stokes and anti-Stokes Raman scattering photons. The fact that spatially separated diamonds were entangled at room temperature is significant because it indicates that  an entangled state can persist in ordinary environments in macroscopic solids, which makes these objects the basis for developing cryogen-free quantum computers. The THz-fast read/write cycle using a diamond optical phonon  amply illustrates the potential of the field". 

We agree with  the rewiever in the part concerning of a practical significance of the results discussed, however a number of details in the  foregoing explanation has to be corrected.

To give the correct explanation we have to take into consideration the following. The authors of \cite{Lee_1} write: "The manifestation of an entangled state is the visibility of the interference pattern of anti-Stokes photons during the registration of coincidences in detecting Stokes and anti-Stokes photons, the visibility being measured by rotating the polarization of these photons". It seems to be correct, however the phenomenon of the interference is fine and subtle phenomenon. Before discusing the results of \cite{Lee_1}, we represent the comment, concerning the   given phenomenon, which is given  in \cite{PC}. Please, the quotations from the given work: "There were revealed very interesting properties of the light flux in the experiments on the interference and diffracion of light described in \cite{Dempster}. It has been found that at very low intensities an interference pattern is not appeared. At the same time the atoms of free silver are appeared on a photographic  plate in the place of photon falling. They represent themselves an embryo, which is much smaller than a light wave length. The particle properties of the light show up in the birth of free silver atoms on a photographic  plate, one by one.  At the same time, when the light intensity is rather large an   interference pattern has been shown up.

The results described in \cite{Dempster} is the good example for the display of the reality of rather complicated structure of EM-field, the Fermi liquid model of which is recently proposed in \cite{DAA}, and it has been shown, that the display of the corpuscular or wave nature of EM-field will
be dependent on experimental conditions. The experimental results on the interference and diffracion of light reported in \cite{Dempster}  seem to be the excellent confirmation for the given conclusion. The observation of the only corpuscular properties at a low light intensity is easily explained by rectlinear propagation of corpuscles - photons, the size of which seems to be not exceeding the very small size of the silver embryo. At the same time, the density of massless boson-atoms of quantized EM-field is rather low  in interslit space at  a low light intensity and the formation of Bloch-like waves do not take place. In fact, the given experiment indicates that there is threshold in the massless boson-atom density for the formation of Bloch-like waves and, correspondingly,  the threshold in light intensities. Consequently, the wave properties of the light can be observed the only by the intensities exceeding the given threshold. In other words owing to the ability of a quantum  Fermi liquid (like to any liquid) to spreading the infill of all interslit space takes place, however the concentration of masless bosons has to be sufficient to realize the interslit space infill. It is interesting to remark, that the representation of EM-field to be quantum liquid is well agree with  a classical representation of the propagation process of the light through small apertures, in which  the apertures are postulated being to be new light point sources. It is  quite similar to  spreading of any liquid through small apertures by the presence of  the pressure [the presence of light pressure is taken into account], for instance, like to spreading of a eau-de-Cologne from a bottle of eau-de-Cologne with a pulverizer in hairdressing saloons.

Therefore, the experimental results of  Dempster and Batho seem to be the most striking argument in the favour of the quantum  Fermi liquid model of EM-field, proposed in 
\cite{DAA}.

We have to remark that all existing in quantum optics theories do not  explain 
 correctly the phenomenon of the interference, in particular, the classical experiment of Young with two slits. According to \cite{Scully}, the appearance of the interference in  Young experiment depend on the coherence degree of two beams of light only  and do not depend on their intensity in contradiction with the results of \cite{Dempster}. All the more, the attempt to consider the propagation of a single photon through two slits simultaneously undertaken by some authors seems to be the grossest blunder".
The foregoing comment is in fact the proof, that the interpretation, proposed in \cite{Lee_1} is incorrect. Really it is based on the part of  Schr\"{o}dinger's theory, describing the only corpuscular aspect [moreover, its probabilistic variant, the possibility to its use has to be argued by authors by  a description of processes characterising by very short times]. The given aspect never can explain   the phenomenon of the  interference, since the photon, although it represent itself an oscillating object and its wave function has  the temporal factor $\exp[(\frac{2\pi E_lt}{h} + \theta_l)i]$,  will be absorbed by an interaction with matter [or reflected]. The boson-atomic structure of EM-field itself is responsible for the phenomenon of the  interference. Let us remember that the existence of analogous structure  for the fields associated with other elementary particles, determining the dynamics of atomic systems, is taken into account in phenomenological Schr\"{o}dinger's theory in implicit form by the equation (\ref{eq10}). For the case of an electomagnetic field the given equation can be represented in the form, which can be deduced from the following picture. The discrete boson-atomic structure of EM-field can geometrically be represented by the superposition of  $N$ onedimensional space-discrete chain lattices - rays [we suggest  that spacing between massless bosons is dependent discretely from the photon energy, propagating along the given chains]. The direction of the rays is coinciding with  a propagation direction. Then the operator of a translation $\mathcal{T}^{m_j}(\vec{r}, \vec{a}_1, \vec{a}_2, ..., \vec{a}_Q)$ for the given lattice set, consisting of $Q$ sublattices can be represented in the form 
\begin{equation}
\label{eq30}
\begin{split}
\mathcal{T}^{m_j}(\vec{r}, \vec{a}_1, \vec{a}_2, ..., \vec{a}_Q) = \sum^Q_{i=1} [\vec{r} + m_j\vec{a}_i].
\end{split}
\end{equation}
Then 
\begin{equation}
\label{eq31}
\begin{split}
\mathcal{T}^{m_j}(\vec{r}, \vec{a}_1, \vec{a}_2, ..., \vec{a}_Q)\sum^Q_{i=1} A_i(\vec{r})= \sum^Q_{i=1} A_i[\vec{r} + m_j\vec{a}_i].
\end{split}
\end{equation}
Since the group of the translations is cyclic abelian group, it has the only
one-dimensional representations. Consequently, we can represent the right side of the relation (\ref{eq31}) in the form
\begin{equation}
\label{eq32}
\begin{split}
\mathcal{T}^{m_j}(\vec{r}, \vec{a}_1, \vec{a}_2, ..., \vec{a}_Q)\sum^Q_{i=1} A_i(\vec{r})= \sum^Q_{i=1} c^m_j A_i(\vec{r}).
\end{split}
\end{equation}
Then, if the $j$-th sublattice has $N_j$ equidistant nodes and if
\begin{equation}
\label{eq33}
\begin{split}
\mathcal{T}^{1_j}(\vec{r}, \vec{a}_1, \vec{a}_2, ..., \vec{a}_Q)\sum^Q_{i=1} A_i(\vec{r})= \sum^Q_{i=1} c^1_j A_i(\vec{r}).
\end{split}
\end{equation}
we will have
\begin{equation}
\label{eq34}
\begin{split}
\mathcal{T}^{N_j}(\vec{r}, \vec{a}_1, \vec{a}_2, ..., \vec{a}_Q)\sum^Q_{i=1} A_i(\vec{r})= \sum^Q_{i=1} (c^1_i)^{N_j} A_i(\vec{r}).
\end{split}
\end{equation}
It is reasonable to suggest that  Born-Carman boundary conditions are taking place, then we will have the following set
\begin{equation}
\label{eq35}
\begin{split}
 \{(c^1_j)^{N_j} = 1\}, j = \overline{1, Q}; 
\end{split}
\end{equation}
Consequently
then we will have the following set
\begin{equation}
\label{eq36}
\begin{split}
 \{(c^1_j) = \exp{\frac{2\pi i\xi_j}{N_j}}\}, j = \overline{1, Q}, 
\end{split}
\end{equation}
The corresponding set of functions, satisfying to the conditions (\ref{eq36}) is
\begin{equation}
\label{eq37}
\begin{split}
\{ A^{\vec{k}_j}_i(\vec{r})= u_{\vec{k}_j}(\vec{r})\exp{\vec{k}_j\vec{r}}\}, i,j = \overline{1, Q},
\end{split}
\end{equation}
if the functions $u_{\vec{k}_j}(\vec{r})$, $j = \overline{1, Q}$, have the period of the sublattice $j$  and the following relation takes place
\begin{equation}
\label{eq38}
\begin{split}
\vec{k}_j N_j = \xi_j \vec{a}^*_j,  j = \overline{1, Q}, \xi_j \in N
\end{split}
\end{equation}   
where $\vec{a}^*_j$ is the vector of the sublattice, being to be reciprocal to the sublattice $j$. Therefore, representing the function $A(x,y,z)\equiv A(\vec{r})$ in the form $A(\vec{r})=\sum^Q_{i=1} A_i(\vec{r})$ the relation (\ref{eq10}) can be rewritten
\begin{equation}
\label{eq39}
\begin{split}
&\psi(x,y,z,t) \equiv \psi(\vec{r}) =\\
&\sum^Q_{i=1}\{u_{\vec{k}_j}(\vec{r})\exp{\vec{k}_j\vec{r}}\sin[{-2\pi\nu_j t}+\frac{S_j(\vec{r})}{K_j}]\},
\end{split}\end{equation} 
where $K_j$ is given by the value $h/2W_j$ and $\nu_j = \frac{E_j}{2\pi K_j}$.
The relation (\ref{eq39}) indicates that the wave function $\psi(\vec{r})$ represents itself the set of  Bloch-like waves. Just the given function allows to describe correctly the wave properties of the light, including the interference. 

Thus to describe correctly the corpuscular-wave dualism it is necessary to use the full variant of Schr\"{o}dinger's theory, taking into consideration two scalar functions. The wave function $\psi(\vec{r})$ is responsible for the wave aspect in a dynamics  of atomic systems. In the case of dynamics studies with the participation of the elementary particles - photons it has the view, given by the relation (\ref{eq39}). The relation (\ref{eq39})
is based both on Schr\"{o}dinger's theory and on the theory considering EM-field, being to be quantum 1D Fermi liquid, theory of which is developed in \cite{DAA}. It seems to be reasonable to suggest, that the relations, analogous to (\ref{eq39}) are correct for the fields, associated with other  elementary particles, incoming in the structures of atomic systems.  

The field scalar function  $\Psi(q_1,q_2,...,q_N)$ above considered is responsible for the description  of corpuscular properties of atomic systems. The concrete analytical expression for the given function is obtained by means of the solution  of the corresponding Schr\"{o}dinger equation.
It is clear,
taking into account the resemblance of the laws of quantum mechanics and both geometrical and undulatory optics, that just  the electromagnetic forces are determining forces in the dynamics of quantum mechanics systems.
The fact is that the dynamics of classical mechanics systems is determined by taking into account the gravitation field.  At the same time, the role of gravitation field for the dynamics of elementary particles with rather small masses seems to be too little, up to have substantial effect in the dynamics of quantum mechanics systems. 

From aforegoing consideration, the direction for the magistral development of the quantum mechanics is crystallized. It is the way of a quest for the description of a single event, which can be realised bu using, in particular, the interpretation of field scalar function, proposed by  Schr\"{o}dinger himself, that is, by one of the main creators of quantum theory.

The comprehension of the origin of the corpuscular-wave dualism indicates the way for the elaboration of the correct quantum  theory of wave phenomena  - interferention, diffraction and so on, which in fact has to be anew created.

Let us give the conclusion on the status of the second main postulate. Its formulation in all textbooks has to be represented in the form of statement, but not a postulate, since the hypothesis of  Schr\"{o}dinger on the existance of the field scalar function, being to be observable quantity, just charge density, is strictly proved for the case of EM-field, the role of which is decisive for the dynamics of the atomic systems. Moreover,  it is indeed describes the state of the system, since the full set of the functions for the description of EM-field is consisting of four scalar functions or equivalently, from one scalar and one vector functions. The information, which is given by an observable vector function of EM-field - field strenth function - is included in an implicit form in Schr\"{o}dinger equation for the field scalar function, that is, it is taking into consideration too. Therefore, the second main postulate in Schr\"{o}dinger formulation [more strictly, his name according to Schr\"{o}dinger was hypothese] can be mathematically strictly grounded, but in popular form aforecited - not. The probabilistic theatise, proposed by Born \cite{Born} is true in special cases, quite correctly indicated by Dirac \cite{Dirac_2}, althogh the given cases embrace the wide range of quantum physics phenomena.

For the explanation of the results of  \cite{Lee_1} we have to take also into consideration the recent paper \cite{CAF}. The concept of  a classical free acoustic  field and its quantization in the cavity were theoretically considered. The equations of the free acoustic field were derived. They coincide in the mathematical form with the first two Maxwell equations for the free electromagnetic  field. In other words, it was shown, that phonon, like to all elementary particles has the own associated field. In  given meaning, bijective mapping betweem photon and phonon systems, used in \cite{Lee_1} seems to be correct, since it is restricted the only by the  consideration of transverse phonons. At the same time  corresponding mapping between photons and longitudinal phonons seems to be absent, since the pair equations for the acoustic field analogous to the second pair of Maxwell equations 
 is absent.  However, it is significant the following. Along with single phonon generation will always present the corresponding acoustic field. Just the only acoustic field, but not phonon itself can belong to two diamond crystals simultaneously and its presence is sufficient to explain correctly the observed results.  There are the following arguments, indicating that the value of the phonon size is  setting too high in more than 14 orders of value in \cite{Lee_1}. There seems to be absent the correct theory, which evaluates the size and gives the representation on the geometical shape of the phonon. At the same time the evaluation can be obtained by the comparison with Su-Schrieffer-Heeger theory of organic conductors \cite{SSH}.  In other words, it is reasonably  to suggest, that phonon is also topological defect like to 
topological defects in Su-Schrieffer-Heeger theory and their sizes then can be comparable. The theoretical value of the coherence length for topological solitons in \textit{trans}-polyacetylene is 7$a$ [$a$ is interatomic spacing in \textit{trans}-polyacetylene carbon chain], and it
is the low boundary in the range $7a-11a$, obtained for  the soliton coherence length from experiments.

 The second example is concerned of a nonapplicability of the probability notion for the correct interpretation of the  experimental results. It is the work \cite{Lee_2}

To demonstrate the quantum
nature of the phonons in diamond, the authors of \cite{Lee_2} "begin by implementing a
quantum memory by means of off-resonant Raman scattering, in
which an optical phonon is generated by Stokes scattering from a
write pulse, with retrieval of the stored excitation stimulated by anti-
Stokes scattering from a subsequent read pulse. Although the
memory storage time $T \approx 7$ ps is too short for applications in longdistance
quantum communications, the duration $\tau \approx 60$ fs of the
read/write pulses is still substantially shorter, giving the memory an
extremely large time-bandwidth product $B \approx T/t \geq 100$.
The Raman interaction in diamond is strong, and, because of the
extremely large bandgap, has near-uniform strength at all optical
wavelengths". "To interpret the experiments", - the authors of \cite{Lee_2} continue - "it is convenient to
adopt a simplified picture of the diamond crystal."They consider the diamond crystal being to be a three-level
$\Lambda$ system comprising the crystal ground state $|0\rangle$, a storage state
with an excited optical phonon $|1\rangle$ and a far off-resonant intermediate
state $|Exciton\rangle$ (strictly a manifold of exciton states in the conduction
band)". It seems to be a bold step on the road to description of macroobjects from quantum-mechanically position, however it seems to be correct. the authors of \cite{Lee_2}report on details foran implementation a diamond
quantum memory,theuy are the following: "A 60 fs write pulse (bandwidth, $\sim 8$ THz) at 808 nm is focused into a 0.5-mm-thick diamond crystal, followed, after a variable delay of $1 \leq T \leq 11$ ps, by a
second, cross-polarized read pulse. Raman scattered light is collected,
frequency and polarization filtered, and fibre-coupled to
photon-counting detectors. If a Stokes photon is detected
from the write pulse, this heralds the creation of an optical
phonon excitation: the diamond memory is now charged. After a
programmable delay $T$, the read pulse retrieves the stored excitation,
mapping it into an anti-Stokes photon".

Especially interesting seems to be the attempt to use the probabilistic version of the interpretation of quantum mechanics for the description of the very interesting experimental results. The authors of \cite{Lee_2} write:
"To characterize the strength of the memory interaction, we
measure the joint probability $P_{S,AS}$ of detecting both a heralding
Stokes photon from the write pulse and an anti-Stokes photon
from the read pulse, and the 'single rates' $P_{S}\approx 9\times 10{-4}$
and $P_{AS} \approx 3\times 10{-6}$, which are the unconditional probabilities
for detecting Stokes or anti-Stokes photons scattered from the
write and read pulses, respectively. The combined collection and
detection efficiency $\eta$ is estimated to be 0.1 percents for both Stokes and
anti-Stokes channels. This suggests we are operating in the spontaneous
scattering regime, with a scattering rate of $P_{S}/\eta \approx 0.1$
Stokes photons per pulse. Further support for this
conclusion comes from our observation of linear scaling of
$P_{S}$ with write/read power, and insensitivity of the read pulse
Stokes scattering to the presence of the write pulse. The probability
that the write pulse generates more than one phonon is
therefore small.
"To verify the non-classical nature of the state of the diamond
crystal during storage",- the authors of \cite{Lee_2} write,- 
"we evaluate the normalized cross-correlation
of the Stokes and anti-Stokes fields, given by $g^{(2)}_{S,AS} = \frac {P_{S,AS}}{P_{S}P_{AS}}$.
Classically, the cross-correlation is upper bounded by the Cauchy-Schwarz inequality $g^{(2)}_{S,AS} \leq \sqrt {g^{(2)}_{S,S}g^{(2)}_{AS,AS}}$, where the autocorrelation
functions of the Stokes and anti-Stokes fields appear
on the right-hand side. The Raman scattered fields are thermal
with autocorrelation functions $g^{(2)}_{S,S} =
g^{(2)}_{AS,AS} = 2$, so measured
values of $g^{(2)}_{S,AS}$
 exceeding 2 are indicative of a true quantum
memory". And further: "For  $T = 1$ ps we observe values up to
$g^{(2)}_{S,AS} = 5.1$, establishing that non-classical states are being stored
and retrieved from the optical phonon modes in our
diamond crystal". The resulting conclusion, that non-classical states are being stored
and retrieved from the optical phonon modes seems to be correct. At the same time the results obtained seem to experimental proof of the inapplicability of the notion of  the probability for the case of  $T = 1$ ps, confirming the remark in \cite{Optics_Communications}.  Really, the system is in mixed state, consisting of three substates and it is oscillates between them. The number of events m, when the system is in the storage state $|1\rangle$, divided on  a full number of events $n$ gives the only the frequency of events, but not the probability [to obtain the probability, $n$ has to be $\rightarrow \infty$]. The violation of the Cauchy-Schwarz inequality for the correlation
functions  seems to be the direct proof that we  are really  dealing with  the frequency of events, but not with the probability at  the delay time $T = 1$ ps. The given interpretation is confirmed by that, that at the delay time $T = 5-6$ ps, which is also 
shorter, than the coherence lifetime of the storage state $|1\rangle$, evaluated to $\approx 7$ ps, the system studied is obeing to the Cauchy-Schwarz inequality, that follows from the Figure 3b in \cite{Lee_2}, indicating, that by the given delay times, the notion of the probability can be used  and the analysis of the authors is correct in this case.

It seems to be understandable, that in order to obtain  a new information on the properties of the system studied and the similar systems, the version of quantum mechanics which can describe the single events has to be used.  Therefore,  the experiment above described indicates the direction of the development of the 
 quantum mechanics theory.  In the given sense we see, that there is a  more great advance
in the development of  a quantum physics experiment in the comparison with the development of the theory.

It was aforegoing concluded, that by the representation of all physical
quantities, which determine  the dynamics of classical mechanics systems  by one of the variants of the matrix
representation of complex numbers we obtain the representation, appropriate for  a description  of quantum-mechanical objects automatically, that is the  new quantum theory, at that, it has been shown  \cite{A_Dovlatova_D_Yerchuck}, \cite{Dmitri_Alla_Andrey} that there is   the infinite number of the variants of the creation of quantum theory. The recent work
\cite{Dodin} is the brilliant confirmation of the given conclusion. The given work is in fact a new original version of quantum mechanics. The field theory methods were used by its building, based on the synthesis of ideas of  a quantum theory construction, which were used by the fathers-creators of the 
quantum theory, existing at present. On the first side, the idea of  Dirac \cite{P.Dirac} of the representation of the field by an ensemble of oscillators was used, however it was treated mathematically in a quite other way. On the second side, the idea of  Heisenberg, Born  and Jordan \cite{Born_Jordan},\cite{Born_Heisenberg} of matrix representation of observables was also used, however a quite other set of observables was choosed. Especially interesting was used the idea of Schr\"{o}dinger \cite{Schroedinger1}, \cite{Schroedinger2},  \cite{Schroedinger3}, \cite{Schroedinger4}, \cite{Schroedinger5}, \cite{Schroedinger6}, \cite{Schroedinger7} above worded of a description
of corpuscular properties of atomic systems, being to be systems of fields, at that corpuscles, which are formed in given fields are elementary particles incoming in atomic structures [naturally with corresponding associated fields]. The optical analogue was used like  to  Schr\"{o}dinger, however  a mathematical  description has been done in another way.

 The only action is not quantized in the paper cited. Its quantization is nevertheless very natural within the new formalism and
can be performed exactly like to the first quantization in the traditional
quantum mechanics. 
A number  of advantages has the quantum mechanics version proposed in \cite{Dodin} in comparison with the traditional
quantum mechanics. Especially significant, that both
 the spaces $\Psi$ and $X$, even more generally, space-time, and commutation relations
need to be postulated, in distinction from the traditional
quantum mechanics.

\section{Conclusions}

The conclusion on the status of the second main postulate is given. Its formulation in all textbooks has to be represented in the form of statement, but not a postulate, since the hypothesis of  Schr\"{o}dinger on the existance of the field scalar function, being to be observable quantity, just charge density, is strictly proved for the case of EM-field, the role of which is argued to be decisive for the dynamics of the atomic systems. Moreover,  it is shown, that it actually describes the state of the system, since the full set of the functions for the description of EM-field is consisting of four scalar functions or equivalently, from one scalar and one vector functions. The information, which is given by an observable vector function of EM-field - field strenth function - is included in an implicit form in Schr\"{o}dinger equation for the field scalar function, that is, it is taking into consideration too. Therefore, the second main postulate in Schr\"{o}dinger formulation [more strictly, his name according to Schr\"{o}dinger was hypothese] can be mathematically strictly grounded, but in popular probabilistic form used in modern textbooks on quantum theory it cannot be proved. The probabilistic theatise, proposed by Born \cite{Born} is true in a number of special cases, quite correctly indicated by Dirac \cite{Dirac_2}. At the same time the given cases embrace the wide range of quantum physics phenomena.

The possible ways of the development of quantum theory are analysed.

\end{document}